\documentclass[sigconf,authorversion=true]{acmart}
\usepackage[utf8]{inputenc}
\usepackage{hyperref}
\usepackage{caption}
\usepackage{subcaption}
\usepackage{tabulary}
\usepackage{textcomp }

\newcommand{\keyword}[1]{\textit{#1}}

\newcommand{\skl}[0]{scikit-learn}

\newcommand{\qq}[0]{\textquotesingle\textquotesingle}
\newcommand{\str}[1]{\qq{}{#1}\qq{}}

\title{Improving the Learnability of Machine Learning APIs by Semi-Automated API Wrapping}

\author{Lars Reimann, Günter Kniesel-Wünsche}
\email{{reimann, gk}@cs.uni-bonn.de}
\affiliation{%
  \institution{Smart Data Analytics, Institute for Computer Science III, University of Bonn}
  \city{Bonn}
  \country{Germany}
}
\date{September 2021}

\copyrightyear{2022} 
\acmYear{2022} 
\setcopyright{acmcopyright}\acmConference[ICSE-NIER'22]{New Ideas and Emerging Results }{May 21--29, 2022}{Pittsburgh, PA, USA}
\acmBooktitle{New Ideas and Emerging Results (ICSE-NIER'22), May 21--29, 2022, Pittsburgh, PA, USA}
\acmPrice{15.00}
\acmDOI{10.1145/3510455.3512789}
\acmISBN{978-1-4503-9224-2/22/05}

\begin{document}

\begin{abstract}
A major hurdle for students and professional software developers who want to enter the world of machine learning (ML), is mastering not just the scientific background but also the available ML APIs. Therefore, we address the challenge of creating APIs that are easy to learn and use, especially by novices. 
However, it is not clear how this can be achieved without compromising expressiveness. We investigate this problem for \skl{}, a widely used ML API. In this paper, we analyze its use by the Kaggle community, identifying unused and apparently useless parts of the API that can be eliminated without affecting client programs. In addition, we discuss  usability issues in the remaining parts, propose related design improvements and show how they can be implemented by semi-automated wrapping of the existing third-party API. 

\end{abstract}



\maketitle 

\section{Introduction}
The success of machine learning (ML) is leading to an ever increasing demand for data scientists. However, software developers who try to fill this gap and have already delved into the necessary scientific ML background still face the challenge of learning how to correctly use the related ML APIs. 
%
The importance of API learnability and usability has been emphasized in many influential studies \cite{Robillard:IEEESOftware2009,Robillard:ESE2011,SouzaCleidsonBentolila:ICSE2009,WangGodfrey:MSR2013,ZibranEtAlWCRE2011}.
Therefore, we aimed to find out whether and how the learning curve can be flattened and correct use 
can be fostered.

Using \skl{} \cite{scikit-learn}, a widely-adopted ML library as a case study, we address the following research questions:

\begin{description}
    \item [RQ1] Can the complexity of the API be reduced without affecting its existing users?
    
    \item [RQ2] Which deviations from established principles of good API design make learning and correct use of ML APIs hard, even for developers who have the required mathematical and software engineering skills?

    \item [RQ3] Can learnability and correct use be improved by a different design of the API?
    
    \item [RQ4] Given an improved API design, can an implementation be derived and maintained 
    automatically, based on the official library? What precisely can be automated and how much human effort is still needed?
\end{description}

 


\section{API Complexity Reduction}
\label{sec:api_complexity}
Existing ML APIs, such as \skl{}, follow a ``one size fits all'' approach, irrespective of user expertise. We questioned this approach, assuming that novices need just a subset of the API and are hence distracted and confused by additional functionalities \cite{ZibranEtAlWCRE2011}. We consider novices to be persons who aim to develop an ML application and have the required mathematical and software engineering background, but no experience using the particular ML API.

\subsection{Empirical Results}
\label{sec:empirical}
\begin{figure}
     \centering
     \begin{subfigure}[b]{0.45\textwidth}
         \centering
         \includegraphics[width=\textwidth]{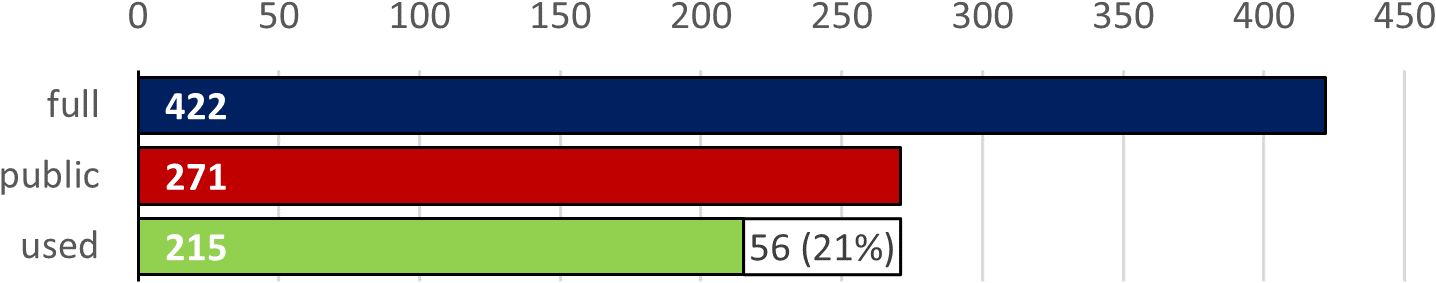}
         \caption{Classes}
         \label{fig:scikit-statistics-bar-chart-classes}
     \end{subfigure}
     \hfill
     \\[1.5ex] 
     \begin{subfigure}[b]{0.45\textwidth}
         \centering
         \includegraphics[width=\textwidth]{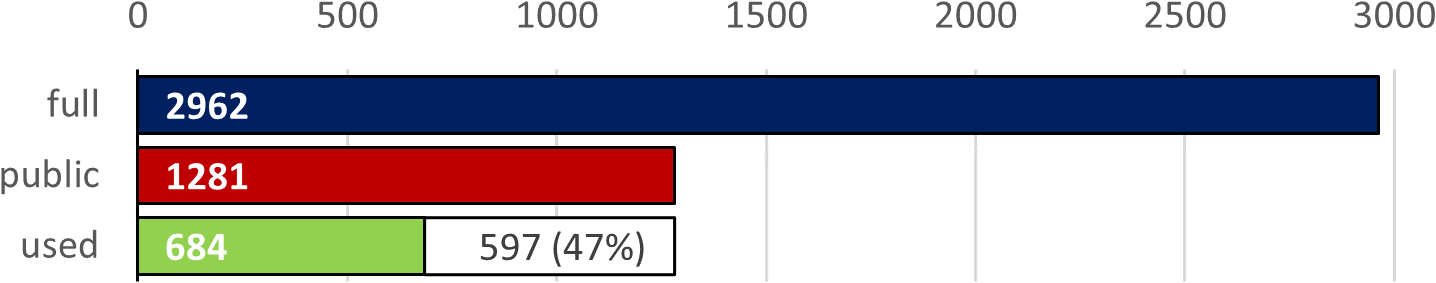}
         \caption{Functions}
         \label{fig:scikit-statistics-bar-chart-functions}
     \end{subfigure}
     \hfill
     \\[1.5ex]
     \begin{subfigure}[b]{0.45\textwidth}
         \centering
         \includegraphics[width=\textwidth]{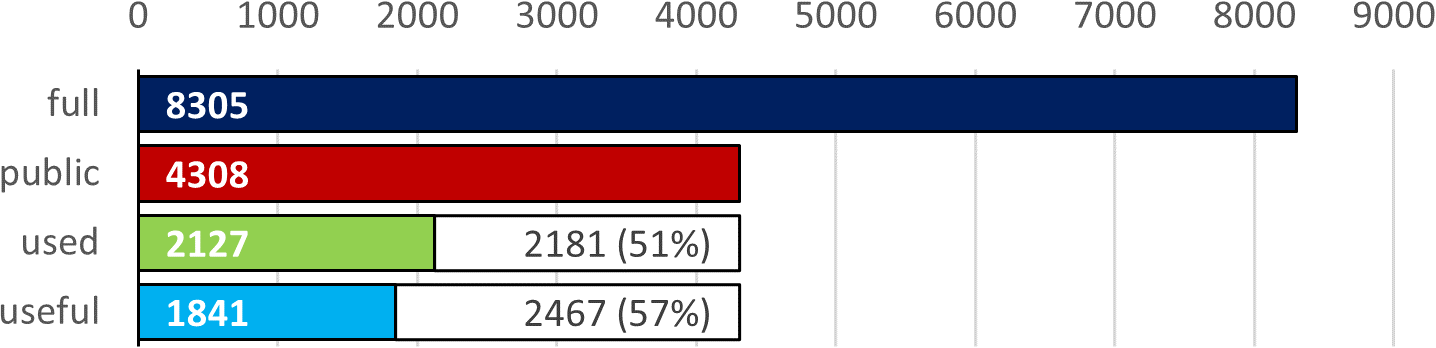}
         \caption{Parameters}
         \label{fig:scikit-statistics-bar-chart-parameters}
     \end{subfigure}
    \vspace{-0.5em}
    \caption{Numbers of classes, functions, and parameters in \skl{}: Dark blue for the entire API, red for public elements, green for elements used in Kaggle competitions, light blue for 
    parameters that are set to at least two different values. The white area shows absolute and relative reductions compared to the public API.}
    \vspace{-1em}
    \label{fig:scikit-statistics-bar-chart}
\end{figure}

To verify our assumption, 
we aimed to quantify the amount of functionalities
that novice users do not need. However, lacking a representative pool of programs written only by novices we started with a broader analysis of all usages that we could find. We did this in two steps:
\begin{description} 
	\item [API statistics] 
	
	    We counted the number of API elements
	    in \skl{} version 0.24.2.
	    In Fig. \ref{fig:scikit-statistics-bar-chart}, the dark blue bars indicate the \emph{total} total numbers of classes, functions, and parameters, whereas the red bars show the respective numbers of \emph{public} elements.   
	    
	\item [Usage statistics] 
	
	    By analyzing 92,402 programs contributed to Kaggle competitions\footnote{https://www.kaggle.com/competitions} by users of any skill level 
	    we found 
	    41,867 that used scikit-learn. Within this pool, we counted the number of \emph{uses} of each class, function and function parameter.
%
%
%
	    Based on this raw data, we computed the additional two bars shown in Fig. \ref{fig:scikit-statistics-bar-chart}. The green bars indicate the number of classes, functions and parameters actually \emph{used} in Kaggle competitions. The light blue bar indicates \emph{useful} parameters, which are set to at least two different values. We call parameters that are always set to the same value \emph{useless}, since they can be replaced by a local variable in the containing function and eliminated from the API.
	    An example of a useless parameter is \textsl{copy\_X} (optional parameter with default \textsl{True}) in the \textsl{Lasso} model, since it is set to \textsl{True} in all 748 cases the constructor is invoked. However, the parameter is not unused since users passed the value \textsl{True} three times explicitly.
\end{description}

From Fig. \ref{fig:scikit-statistics-bar-chart} we can infer the reduction of the API's complexity if users were presented just with the used classes and functions and useful parameters. This benefits novices and experts alike:
%
%
%
%
%
%
It shows that even among the public elements, 56 classes (21\%) and 597 functions (47\%) are unused, while 2467 parameters (57\%) are useless. These API elements can be eliminated from the user-visible API without affecting the functionality of the analysed applications. The improvements are huge and can be a significant time-saver when learning a new API.

\subsection{Plausibility Check}
\label{sec:PlausibilityCheck}
We wanted to be sure that the above results are not biased by the choice of Kaggle competitions as sample data, but can be generalized to other ML applications.
%
To this end, we inspected unused classes and functions, and useless parameters, in search of reasons why elements are generally not needed for application development.
\begin{description}
    \item [Unused classes] We found that most unused classes are indeed clearly irrelevant for application programmers, being designed just for extending the API by additional models and internal code reuse. Examples include mixins, base classes, and a few unused, exotic ML models, like \textsl{MultiTaskLasso}.   
    
    \item  [Unused functions] Similarly, several groups of unused functions are intended for API extensions only, such as validation functions for inputs. Some are related to sample datasets (\textsl{fetch\_california\_housing}), which are clearly not used in applications that include their own data. 
    However, there are also unused core ML functions, often composite ones (\textsl{fit\_transform}, \textsl{fit\_predict} etc.) that exist for API consistency, and which we do not want to remove for this reason. We guess they are not called because users might  call the component functions separately (e.g. \textsl{fit} or \textsl{transform}). Finally, some introspection functions (\textsl{decision\_functions}) are not used either and also deserve further investigation. 
    
    \item [Unused and useless parameters] Given the sheer number of parameters, we have not yet completed categorizing them. However, our findings to date confirm the generality of the empirical results, also for parameters.
    
\end{description}

Since our plausibility check suggests that the results obtained for Kaggle can be generalized, we conclude that for \skl{}, the complexity of the API can indeed be reduced without affecting existing users (RQ1) by roughly the numbers indicated in 
Fig. \ref{fig:scikit-statistics-bar-chart}, by eliminating unused and useless elements from the public API.

\section{API Redesign}
\label{sec:api-redesign}

The manual inspection done for our plausibility check 
revealed aspects that were not evident from the statistical data.
We found that even the \emph{used and useful} part of the API is subject to design and usability problems that deserve specific attention. This motivated our research questions RQ2 and RQ3. In the following, we address these questions in turn, enumerating found usability issues 
(RQ2) and proposing related design improvements (RQ3).

\subsection{Proliferation of Primitives}
\label{sec:proliferation}

One of the design principle of \skl{} is \keyword{non-proliferation of classes}, which means classes are only added for learning algorithms. All other concepts are expressed by primitive types, or types from other scientific libraries, like numpy or scipy \cite{sklearn_api}. 
%
%
This, however, results in long lists of parameters of primitive types, contradicting established design principles.
McConnell's guideline, based on psychological research, is to use at most seven parameters \cite[p. 178]{code_complete}. 
Martin (\cite[p. 288]{clean_code}) suggest to keep parameter lists as short as possible and to avoid more than three parameters. 
While the removal of unused parameters (Sec. \ref{sec:empirical}) 
alleviates the problem, it does not eliminate it. This section discusses related issues and ways to address them without sacrificing the generality of the API:

\begin{description} 

    \item [Tangled concerns] \label{item:tangling}
        The long parameter lists of the constructors of ML models hide the important \emph{hyperparameters} between parameters that are unrelated to the core ML concern, such as debugging or performance options.
        For example, the constructor of the \textsl{SVC} model has 15 parameters (all of which are used) for a multitude of concerns: \textsl{kernel} is a hyperparameter, while \textsl{verbose} is for debugging, and \textsl{cache\_size} concerns performance.
        
        Such parameters can be removed from the constructors of ML models leaving only the hyperparameters, drastically increasing their visibility.
        If needed, attributes for non-ML concerns in the created object can anyway be set to non-default values before calling its \textsl{fit} method to start training.
        
        

    \item[Implicit dependencies]
        Flat lists of inter-dependent parameters can allow combinations of parameter values that are semantically wrong. For instance, it is non-obvious that the \textsl{degree} parameter of the SVC constructor should only be set if the \textsl{kernel} parameter is set to \textsl{\str{poly}}. The invocation \textsl{SVC(\allowbreak{}kernel=\allowbreak{}\str{linear},\allowbreak{} degree=\allowbreak{}3)} is possible but most surely an error, since the degree value will be ignored. However, \textsl{SVC(\allowbreak{}kernel=\allowbreak{}\str{poly},\allowbreak{} degree=\allowbreak{}3)} is correct.
        
        Grouping dependent parameters in one object, called a \keyword{parameter object} in \cite{refactoring}, improves understandabilty \cite{LaceraEtAl:JSS2020}, makes dependencies clear, and prevents any accidental misuse of the API. In our example, we create different variants of \textsl{Kernel} objects, with a \textsl{degree} parameter only in the \textsl{\str{poly}} variant. Via \textsl{SVC(\allowbreak{}Kernel.linear)} and  \textsl{SVC(\allowbreak{}Kernel.poly(\allowbreak{}degree=\allowbreak{}3))} we can then create legal SVC configurations, but the erroneous example above will be impossible to create. 
        
    \item[Magic strings]
        Scikit-learn often uses string parameters that only have a limited set of legal values\footnote{
        In 2007, when \skl{} was initially released, Python offered no alternative. Later, the string types had to be kept for backwards compatibility.}. This has the downside that statically ensuring use of a valid value is difficult. This problem is similar to the anti-pattern of using integers to encode such limited value sets \cite{effective_java, clean_code}.
        
        The problem is worsened by the fact that \skl{} sometimes does not validate parameters. For example, the \textsl{criterion} parameter of the \textsl{DecisionTreeClassifier} should either be \textsl{\str{gini}} or \textsl{\str{entropy}}. However, the creation of an instance with the call \textsl{DecisionTreeClassifier(\allowbreak{}criterion=\allowbreak{}\str{giny})} is silently accepted but, when we later call the \textsl{fit} method on the model, the program crashes. The error message \textsl{KeyError: \str{giny}} gives us only a vague hint about the source of the error. 
        
        Using instead \keyword{enums} in conjunction with \keyword{type hints}\footnote{\url{https://docs.python.org/3/library/typing.html} (added in Python 3.5)} lets static analysis tools ensure that a correct value is passed.

        
\end{description} 

The proposed introduction of parameter objects and enums would increase the number of classes, contrary to the non-pro\-li\-fe\-ra\-tion of classes design of \skl{}. It remains to be investigated which design is easier to learn and use correctly. We are confident that, with a careful redesign, (1) the improved error-resilience of the redesigned API is worth the effort and (2) the increase in the size of the API will be outweighed by the usage-based reduction discussed in Sec. \ref{sec:api_complexity}.

\subsection{Module Structure}
\label{sec:moduleStructure}
Module structure is another target for API redesign: 
%
%
Models for supervised learning are grouped by model category, such as Support Vector Machine (SVM) models. The task they solve is only indicated by a suffix on the class name.
For example, \textsl{SVC} is an SVM model for classification, and  \textsl{SVR} is one for regression.
Similarly, metrics for model evaluation are grouped in \textsl{sklearn\allowbreak{}.metrics}, regardless of task, even though the sets of metrics applicable to different tasks are disjoint.


However, developers are interested in fulfilling a \emph{task}, rather than exploring a particular \emph{family of algorithms}.
Therefore, we suggest grouping of models and metrics by task, like in Apache Spark MlLib \cite{mllib}.
This speeds up the search for models applicable to a task, or metrics suited to evaluate their performance, and makes it obvious, which models and metrics can be used interchangeably. The importance of aligning software components  with tasks 
has already been discussed in \cite{kersten:mylar, wang:feature_location, thung:api_method_recommendation}.

\section{API Wrapping}
\label{sec:api-wrapping}

The design improvements discussed so far should not be misunderstood as a proposal for redesigning \skl{}, a high-quality, widely used ML library. Neither do we want to reinvent the wheel, nor break the huge number of programs using \skl{}. 

Our vision is to wrap the existing \skl{} library into an API that implements the suggested improvements so that it is more suited for novices. However, the size of the \skl{} library prohibits manual wrapper creation and maintenance. Thus, our approach raises the two challenges, summarized in RQ4: (1) automated initial wrapper creation and (2) automated update of wrappers, whenever a version of the \skl{} library is released.

\subsection{Initial Wrapper Creation}
\label{sec:wrapper creation}
As a basis for automation, the manually derived design improvements outlined previously need to be expressed in a machine-readable form that specifies precisely which \emph{changes} should be performed on which \emph{API elements}.  

\begin{table}
\centering

\begin{tabulary}{\linewidth}{cL}
 \hline
 Annotation & Meaning \\
 \hline\hline
 \textsl{@remove} & Remove unused \& unnecessary API element. \\ 
 \hline
 \textsl{@attribute} & Remove parameter from the constructor of a model and keep it only as a model attribute. \\
 \hline
 \textsl{@group} & Group dependent parameters as an object. \\
 \hline
 \textsl{@enum} & Replace string type with enum type. \\
 \hline
 \textsl{@move} & Move class / global function to another module. \\
 \hline
\end{tabulary}

\vspace{0.5\baselineskip}
\caption{Annotations that can be set in the annotation editor and their implied effect on wrapper creation.}
\vspace{-2.4em}
\label{tab:annotations}
\end{table}

We do this by annotating elements of the \skl{} API. Each annotation type (Table \ref{tab:annotations}) corresponds to one of the improvements outlined in Sec. \ref{sec:api_complexity} and \ref{sec:api-redesign}. A user can 
attach these annotations to classes, functions and parameters in a web-based annotation editor\footnote{\url{https://github.com/lars-reimann/api-editor}}. To reduce the workload of annotators and guide them to relevant elements, an initial set of \textsl{@remove} annotations
is created automatically from the usage data described in Sec. \ref{sec:empirical}.   
%
Based on the annotations and the original \skl{} API, the new API is inferred and corresponding \keyword{wrappers} \cite{gof} are generated.

\subsection{Wrapper Update}
\label{sec:wrapperUpdate}

The authors of \skl{} follow a strict release policy aimed not to break existing clients. API elements scheduled for removal are deprecated two versions in advance. Renamings and moves are implemented by deprecating the existing element and adding a new one. Similarly, changes of legal parameter values are implemented by deprecating the current value and adding a new one. This addition of new elements and forwarding from old to new ones is basically a wrapping step. 
Thus, a lightweight evolution policy can reflect the deprecations, additions and deletions from \skl{} in our adapted API, delegating the task of updating client programs to users of our API. Additions and deletions can be identified by our code analysis (Sec. \ref{sec:empirical}). Deprecation information can be extracted automatically from the \skl{} documentation using a mix of parsing,
and rudimentary natural language processing (NLP). This avoids having to repeat the initial manual annotation work in any future releases. Only a fraction of the added elements might need new annotations.

\section{Related Work}

\paragraph{API usability}



Among the rich body of literature about API usability, other contributions that were not cited so far but have influenced our work include the empirical study of API usability problems by Piccioni et al. \cite{piccioni2013}, the evaluation of API usability using Human-Computer Interaction methods by Grill et al. \cite{grill2012}, general recommendations for API designers by Myers and Stylos \cite{myers:api_usability}, and the 
JavaDoc alternative by Stylos et al. \cite{myers:api_documentation_usage}, which takes usage data of an API as input to create better documentation.
%
However, we only found one other assessment of APIs in the context of ML, namely by Király et al. \cite{kiraly2021designing}, who compiled the interfaces and design patterns underlying \skl{} and other ML APIs and derived high-level guidelines for ML API design but didn't investigate their practical implications on a concrete framework, such as \skl{}. In addition their scitype concept also makes the notion of tasks explicit, which provides a formal basis for the task-oriented restructuring that we propose in Sec. \ref{sec:moduleStructure}.

\paragraph{ML for novices}  
Our simplified API has been developed as part of the Simple-ML project \cite{Gottschalk2019, guidance}, as a back-end for a user-friendly Domain Specific Language (DSL) for ML. The Simple-ML API and DSL are embedded into an Integrated Development Environment (IDE) for ML workflows. The entire system is tailored to the needs of data science novices, to make ML more accessible.



\paragraph{Wrapper creation} Our API wrapping approach described in Sec. \ref{sec:api-wrapping} is similar to the one of 
Hossny et al. \cite{paas_adapters}, who use semantic annotations to hide  proprietary APIs of different cloud providers behind a generic one, in order to combat vendor lock-in. Wrappers are automatically created so they can easily be kept up-to-date.

\paragraph{API evolution} The vast literature about techniques to tackle the evolution of an API and automatically keep \emph{client} code up-to-date is
compiled in a recent survey \cite{evolution_survey}. However, our focus is not on updating client code but on updating our generated wrappers. For this, we need to be able to detect changes in \skl{} and identify conflicts with respect to the changes implied by our annotations. An extensive review of related change detection and change merging approaches can be found in \cite{mens2002:merging}.



\section{Threats to Validity}
\label{sec:threats-to-validity}

\paragraph{Usage data} For now, we only pulled usage data from Kaggle programs that were linked to competitions. This can skew results, since we got groups of programs that solve the same problem and, therefore, might employ similar methods. Moreover, some competitions were very popular, so we could get 1000 entries\footnote{The maximum that can be retrieved via Kaggle's REST API}, whereas others only had a handful.

\paragraph{\skl{} v1.0} In the meantime, a new version of \skl{} has been released. The few contained API changes\footnote{\url{https://scikit-learn.org/stable/whats_new/v1.0.html}} slightly change the absolute numbers but not the percentages reported in Sec. \ref{sec:api_complexity}.


\section{Future Plans}
\label{sec:future}

To eliminate possible bias (Sec. \ref{sec:threats-to-validity}), we want to extend the gathered Kaggle data with additional \skl{} usage data from GitHub repositories. Afterwards, our plausibility check (Sec. \ref{sec:PlausibilityCheck}) for parameters needs to be completed.
%
The suggestions for API redesign (Sec. \ref{sec:api-redesign}) initially had to be entered manually in the annotation editor. We have started automating this by extracting information from the documentation of \skl{}, using a mix of parsing (for the structured parts of the documentation) and natural language processing (for the rest).
Task-oriented restructuring of modules (Sec. \ref{sec:moduleStructure}) will be partially automated by checking the suffix of model names (e.g Classifier vs. Regressor). The annotation editor (Sec. \ref{sec:api-wrapping}) is fully implemented. Annotation of \skl{} is in progress and the added annotations will be used as input for wrapper generation, and as a test oracle for automated extraction of necessary design changes from the documentation.
%
%
%
Finally, the resulting API needs to be carefully evaluated through usability studies to determine whether it really yields the conjectured improvement.

\section{Conclusion}


In this paper, we have presented a novel approach for easing the learning and correct use of a complex API. Unlike approaches to learnability that focused on improving the documentation of an API \cite{Robillard:ESE2011, Robillard:IEEESOftware2009}, we investigated whether API \emph{complexity} can be reduced, to avoid overwhelming application developers.
Based on analysis of the API's usage in 41,867 programs, we showed that 21\% of classes, 47\% of functions, and 57\% of parameters can be eliminated, without affecting any of the analysed programs (Fig. \ref{fig:scikit-statistics-bar-chart}).

Analysing code and documentation of the remaining API elements, we found 
a proliferation of elements of primitive types, which
impedes learnability and eases API misuse. Tangling of different concerns in long parameter lists hides the essential hyperparameters. Hard to debug errors arise from implicit dependencies of parameters and hidden narrowing of parameter domains in code that behaves correctly just on a subset of the values that can be passed (Sec. \ref{sec:api-redesign}).



We showed that design improvements inspired by refactorings can solve these issues. However, since we cannot refactor a third-party API, 
we proposed an alternative approach: semi-automated API wrapping. We showed how a layer of wrappers 
that implement an improved API can be built and kept up-to-date with minimal manual effort, based on analysis of the original library's source code and documentation, usage statistics, and code generation. 

%
The only ML-specific argument in our discussion is the distinction between hyperparameters and parameters for other concerns. This notion can be generalized to a distinction between domain-specific parameters and others. 
Hence, semi-automated API wrapping should be applicable to other domains and APIs, as a general way to reduce API complexity and improve API design.


\begin{acks}
We want to thank the reviewers for their very constructive comments and inspiring suggestions for future work. This work was partially funded by the German Federal Ministry of Education and Research (BMBF) under project Simple-ML (01IS18054). 
\end{acks}

\newpage

\bibliographystyle{ACM-Reference-Format}
\bibliography{references}

\end{document}